\documentclass[aps, pra, twocolumn,showpacs,preprintnumbers,amsmath,amssymb]{revtex4}

\newcommand{\ket}[1]{\mbox{$ | #1 \rangle $}}
\newcommand{\bra}[1]{\mbox{$ \langle #1 | $}}
\usepackage[dvips]{graphicx}

\begin{document}

\preprint{}

\title{Quantum secure communication protocols based on entanglement swapping}

\author{Jian Wang}

 \email{jwang@nudt.edu.cn}

\affiliation{School of Electronic Science and Engineering,
\\National University of Defense Technology, Changsha, 410073, China }
\author{Quan Zhang}
\affiliation{School of Electronic Science and Engineering,
\\National University of Defense Technology, Changsha, 410073, China }
\author{Chao-jing Tang}
\affiliation{School of Electronic Science and Engineering,
\\National University of Defense Technology, Changsha, 410073, China }


\begin{abstract}
We present a quantum secure direct communication protocol and a
multiparty quantum secret sharing protocol based on
Einstein-Podolsky-Rosen pairs and entanglement swapping. The present
quantum secure direct communication protocol makes use of the ideal
of block transmission. We also point out that the sender can encode
his or her secret message without ensuring the security of the
quantum channel firstly. In the multiparty quantum secret sharing
protocol, the communication parties adopt checking mode or encoding
mode with a certain probability. It is not necessary for the
protocol to perform local unitary operation. In both the protocols,
one party transmits only one photon for each Einstein-Podolsky-Rosen
pair to another party and the security for the transmitting photons
is ensured by selecting $Z$-basis or $X$-basis randomly to measure
the sampling photons.
\end{abstract}

\maketitle


%
%
Quantum cryptography has been one of the most promising applications
of quantum information science. It utilizes quantum effects to
provide unconditionally secure information exchange. Quantum key
distribution (QKD) which provides unconditionally secure key
exchange has progressed quickly \cite{bb84,b92,e91,bbm92}. In recent
years, a good many of other quantum cryptography schemes have also
been proposed and pursued, such as quantum secret sharing
(QSS)\cite{hbb99,kki99,zhang,zhangman,li,jwang1,gg03,zlm05,xldp04},
quantum secure direct communication (QSDC)
\cite{beige,Bostrom,Deng,denglong,cai1,jwang2,cai4,jwang3,jwang4,cw1,cw2,tg,zjz}.
QSS is the generalization of classical secret sharing to quantum
scenario and can share both classical and quantum messages among
sharers. Many works have been carried out in both theoretical and
experimental aspects after the pioneering QSS scheme proposed by M.
Hillery et al. in 1999 (hereafter called HBB99)\cite{hbb99}. We can
classify the QSS schemes as schemes using entanglement or schemes
without entanglement. The HBB99 scheme is based on a three-particle
entangled Greenberger-Horne-Zeilinger (GHZ) state. A. Karlsson et
al. proposed a QSS scheme using two-particle Bell states
\cite{kki99}. Based on multi-particles entangled states and
teleportation, we presented a multiparty QSS scheme of QSDC
\cite{jwang1}. G. P. Guo and G. C. Guo presented a QSS scheme where
only product states are employed \cite{gg03}. Z. J. Zhang et al.
\cite{zlm05} proposed a QSS scheme using single photons. QSDC's
object is to transmit the secret message directly without first
establishing a key to encrypt them, which is different to QKD. QSDC
can be used in some special environments which has been shown by K.
Bostr\"{o}em et al. and F. G. Deng et al.\cite{Bostrom,Deng}. Many
researches have been carried out in QSDC. These works can also be
divided into two types, one utilizes single photons
\cite{denglong,cai1,jwang2}, the other utilizes entangled state
\cite{Bostrom,Deng,cai1,cai4,jwang3,jwang4,cw1,cw2,tg,zjz}. F. G.
Deng et al. proposed a QSDC scheme using batches of single photons
which serves as one-time pad \cite{denglong}. Q. Y. Cai et al.
presented a deterministic secure direct communication scheme using
single qubit in a mixed state \cite{cai1}. We proposed a QSDC scheme
based on order rearrangement of single photons \cite{jwang2}. The
QSDC scheme using entanglement state is certainly the mainstream. K.
Bostr\"{o}em and T. Felbinger proposed a "Ping-Pong" QSDC protocol
which is quasi-secure for secure direct communication if perfect
quantum channel is used \cite{Bostrom}. Q. Y. Cai et al. pointed out
that the "Ping-Pong" Protocol is vulnerable to denial of service
attack or joint horse attack with invisible photon \cite{cai2,cai3}.
They also presented an improved protocol which doubled the capacity
of the "Ping-Pong" protocol \cite{cai4}. F. G. Deng and G. L. Long
put forward a two-step QSDC protocol using Einstein-Podolsky-Rosen
(EPR) pairs \cite{Deng}. We presented a QSDC scheme using EPR pairs
and teleportation \cite{jwang3} and a multiparty controlled QSDC
scheme using GHZ states \cite{jwang4}.

Entanglement swapping can entangle two quantum systems that do not
have direct interaction with each other. It plays an important role
in quantum information. There are also many quantum cryptography
schemes using entanglement swapping. Z. J. Zhang et al. presented a
multiparty QSS scheme \cite{zhangman} and a QSDC scheme based on
entanglement swapping and local unitary operations \cite{zhangman2}.
Y. M. Li put forward a multiparty QSS of quantum information by
swapping quantum entanglement \cite{li}. Based on entanglement
swapping, T. Gao et al. proposed two QSDC schemes using GHZ states
and Bell states, respectively \cite{gao1,gao2}. We first describe
entanglement swapping simply. The four Bell states are
\begin{eqnarray}
\ket{\phi^\pm}=\frac{1}{\sqrt{2}}(\ket{00}\pm\ket{11}),\nonumber\\
\ket{\psi^\pm}=\frac{1}{\sqrt{2}}(\ket{01}\pm\ket{10}).
\end{eqnarray}
Suppose two distant parties, Alice and Bob, share \ket{\phi^+_{12}}
and \ket{\phi^+_{34}} where Alice has qubits 1 and 4, and Bob
possesses 2 and 3. Note that
\begin{eqnarray}
\label{1}
\ket{\phi^+_{12}}\otimes\ket{\phi^+_{34}}&=&\frac{1}{2}(\ket{\phi^+_{14}}\ket{\phi^+_{23}}+\ket{\phi^-_{14}}\ket{\phi^-_{23}}\nonumber\\
&
&+\ket{\psi^+_{14}}\ket{\psi^+_{23}}+\ket{\psi^-_{14}}\ket{\psi^-_{23}}.
\end{eqnarray}
After Bell basis measurement on qubits 1 and 4, the state of the
qubits 1, 2, 3, 4 collapses to \ket{\phi^+_{14}}\ket{\phi^+_{23}},
\ket{\phi^-_{14}}\ket{\phi^-_{23}},
\ket{\psi^+_{14}}\ket{\psi^+_{23}} and
\ket{\psi^-_{14}}\ket{\psi^-_{23}} each with probability 1/4. If
Alice and Bob share other Bell states, similar results can be
achieved. We give the relations used in our paper, as
\begin{eqnarray}
\label{2}
\ket{\phi^-_{12}}\otimes\ket{\phi^+_{34}}&=&\frac{1}{2}(\ket{\phi^+_{14}}\ket{\phi^-_{23}}+\ket{\phi^-_{14}}\ket{\phi^+_{23}}\nonumber\\
&
&+\ket{\psi^-_{14}}\ket{\psi^+_{23}}+\ket{\psi^+_{14}}\ket{\psi^-_{23}},\nonumber\\
\ket{\psi^+_{12}}\otimes\ket{\phi^+_{34}}&=&\frac{1}{2}(\ket{\phi^+_{14}}\ket{\psi^+_{23}}-\ket{\phi^-_{14}}\ket{\psi^-_{23}}\nonumber\\
&
&+\ket{\psi^+_{14}}\ket{\phi^+_{23}}-\ket{\psi^-_{14}}\ket{\phi^-_{23}},\nonumber\\
\ket{\psi^-_{12}}\otimes\ket{\phi^+_{34}}&=&\frac{1}{2}(\ket{\phi^-_{14}}\ket{\psi^+_{23}}-\ket{\phi^+_{14}}\ket{\psi^-_{23}}\nonumber\\
&
&+\ket{\psi^-_{14}}\ket{\phi^+_{23}}-\ket{\psi^+_{14}}\ket{\phi^-_{23}},
\end{eqnarray}

We then present a QSDC protocol using EPR pairs and entanglement
swapping. Suppose the sender Alice wants to transmit her secret
message directly to the receiver Bob.

(1) Alice prepares an ordered $N$ EPR pairs. Each of the EPR pairs
is in the state \ket{\phi^+_{12}}. We denotes the ordered $N$ EPR
pairs with \{[P$_1(1)$,P$_1(2)$], [P$_2(1)$,P$_2(2)$], $\cdots$,
[P$_N(1)$,P$_N(2)$]\}, where the subscript indicates the order of
each EPR pair in the sequence, and 1, 2 represents the two-photon of
each pair. Alice takes one photon from each state to form an ordered
partner photon sequence [P$_1(1)$, P$_2(1)$,$\cdots$, P$_N(1)$],
called $S_1$ sequence. The remaining partner photons compose $S_2$
sequence, [P$_1(2)$, P$_2(2)$,$\cdots$, P$_N(2)$]. Bob also prepares
$N$ EPR pairs each of which is in the state \ket{\phi^+_{34}}. In
the same way, Bob divides the $N$ EPR pairs into $S_3$ sequence,
[P$_1(3)$, P$_2(3)$,$\cdots$, P$_N(4)$] and $S_4$ sequence,
[P$_1(4)$, P$_2(4)$,$\cdots$, P$_N(4)$]. Alice then sends the $S_2$
sequence to Bob. Bob sends the $S_4$ sequence to Alice at the same
time.

(2) After confirming the two parties have received the photon
sequence, Alice selects randomly a sufficiently large subset from
the photon sequence for eavesdropping check. She chooses randomly
one of the two measuring basis $Z$-basis (\ket{0}, \ket{1}) and
$X$-basis
(\ket{+}=$\frac{1}{\sqrt{2}}$(\ket{0}+\ket{1}),\ket{-}=$\frac{1}{\sqrt{2}}$(\ket{0}-\ket{1}))
to measure the photon 1. She then tells Bob the positions of the
sampling photons, the measuring basis she has chosen and her
measurement result. Bob measures the photon 2 in the same measuring
basis as Alice and compares his result with Alice's. He thus can
evaluate the error rate of the transmission of the $S_2$ sequence.
To ensure the security of the transmission of the $S_4$ sequence,
Alice and Bob utilize the same method to check eavesdropping. If the
error rate exceeds the threshold, they abort the protocol.
Otherwise, they continue to the next step.

(3) After ensuring the security of the EPR pairs, Alice encodes her
secret message on the photon 1. She performs one of the four unitary
operations
\begin{eqnarray}
& &I=\ket{0}\bra{0}+\ket{1}\bra{1},\nonumber\\
& &\sigma_x=\ket{0}\bra{1}+\ket{1}\bra{0},\nonumber\\
& &i\sigma_y=\ket{0}\bra{1}-\ket{1}\bra{0},\nonumber\\
& &\sigma_z=\ket{0}\bra{0}-\ket{1}\bra{1}.
\end{eqnarray}
on the photon 1, according to her secret message. The operations
$I$, $\sigma_x$, $i\sigma_y$ and $\sigma_z$ denote secret message
``00'', ``01'', ``10'' and ``11'', respectively. Alice then performs
Bell basis measurement on the photons 1 and 4 and publishes her
measurement results.

(4) Bob measures the photons 2 and 3 in the Bell basis. Because of
the operation $I$ ($\sigma_x$, $i\sigma_y$, $\sigma_z$ ) performed
on the photon 1, the state \ket{\phi^+_{12}} is changed to
\ket{\phi^+_{12}} (\ket{\psi^+_{12}}, \ket{\psi^-_{12}},
\ket{\phi^-_{12}}). Note that the equations \ref{1} and \ref{2}. Bob
can then deduce Alice's secret according to his result and Alice's
result, as illustrated in Tabel 1. For example, if Alice's result is
\ket{\psi^-_{14}} and Bob's result is \ket{\phi^-_{23}}, Alice's
secret must be ``01''. That is to say Alice performed $\sigma_x$
operation on the photon 1.
\begin{table}[h]
\caption{The recovery of Alice's secret message }\label{Tab:one}
  \centering
    \begin{tabular}[b]{|c|c|c| c|} \hline
     Alice's secret & \{Alice's result, Bob's result\}\\ \hline
      \ 00 ($I$) & \{\ket{\phi^+_{14}}, \ket{\phi^+_{23}}\} \{\ket{\phi^-_{14}}, \ket{\phi^-_{23}}\}\\
       & \{\ket{\psi^+_{14}}, \ket{\psi^+_{23}}\} \{\ket{\psi^-_{14}}, \ket{\psi^-_{23}}\}\\ \hline
       \ 01 ($\sigma_x$) & \{\ket{\phi^+_{14}}, \ket{\psi^+_{23}}\} \{\ket{\phi^-_{14}}, \ket{\psi^-_{23}}\}\\
       & \{\ket{\psi^+_{14}}, \ket{\phi^+_{23}}\} \{\ket{\psi^-_{14}}, \ket{\phi^-_{23}}\}\\ \hline
        \ 10 ($i\sigma_y$) & \{\ket{\phi^-_{14}}, \ket{\psi^+_{23}}\} \{\ket{\phi^+_{14}}, \ket{\psi^-_{23}}\}\\
        & \{\ket{\psi^-_{14}}, \ket{\phi^+_{23}}\} \{\ket{\psi^+_{14}}, \ket{\phi^-_{23}}\}\\ \hline
         \ 11 ($\sigma_z$) & \{\ket{\phi^+_{14}}, \ket{\phi^-_{23}}\} \{\ket{\phi^-_{14}}, \ket{\phi^+_{23}}\}\\
         & \{\ket{\psi^-_{14}}, \ket{\psi^+_{23}}\} \{\ket{\psi^+_{14}}, \ket{\psi^-_{23}}\}\\ \hline
        \end{tabular}
\end{table}

The security for the present protocol is the same as that for BBM92
protocol. To ensure the security of the transmission of the $S_2$
and $S_4$ sequence, the communication parties perform $Z$-basis or
$X$-Basis measurements on the sampling photons, which is similar to
that of BBM92 protocol. Only after confirming the security of the
quantum channel could Alice encode her secret message on the photon
1 and announce her measurement result. Thus our protocol is
unconditional secure.

Actually, in the present QSDC scheme, Alice can encode her secret
message directly on the EPR pairs without ensuring the security of
the quantum channel firstly. Alice prepares a batch of EPR pairs
each of which is in one of the four Bell states according to her
secret message. The states \ket{\phi^+}, \ket{\phi^-}, \ket{\psi^+}
and \ket{\psi^-} represent the secret message ``00'', ``01'', ``10''
and ``11'', respectively. Alice inserts randomly some sampling EPR
pairs in the encoding sequence for eavesdropping check. Similar to
the step 2 in the present protocol, Alice and Bob chooses randomly
$Z$-basis or $X$-basis to measure the sampling photons. Only after
confirming the security of the quantum channel could Alice publish
her results of Bell basis measurement. After obtaining Alice's
results, Bob can then recover her secret message.

We then present a multiparty QSS protocol using EPR pairs and
entanglement swapping. We first present a three-party QSS protocol
and then generalize it to a multiparty QSS one. Suppose Alice want
to share a random key with Bob and Charlie.

(1) Alice, Bob and Charlie agree that the four Bell states
\ket{\phi^+}, \ket{\phi^-}, \ket{\psi^+} and \ket{\psi^-} represent
the two-bit information ``00'', ``01'', ``10'' and ``11'',
respectively.

(2) Alice (Bob, Charlie) prepares an EPR pair in the state
\ket{\phi^+_{12}} (\ket{\phi^+_{34}}, \ket{\phi^+_{56}}). Alice
(Bob, Charlie) send the photon 2 (4, 6) to Bob (Charlie, Alice).

(3) Alice chooses one of the two modes, checking mode with
probability $p$ and encoding mode with probability $1-p$. If Alice
selects checking mode, the procedure goes to (C4), otherwise they
perform the step (E4).

(C4) Bob performs $Z$-basis or $X$-basis measurement randomly on the
photon 2 and informs Alice the measuring basis he has chosen and his
measurement result. Alice then measures the photon 1 in the same
measuring basis as Bob and compares her result with Bob's. The
method of eavesdropping check is similar to that of BBM92 protocol,
which ensures the security of the transmission of the photon 2. If
there is no eavesdropper, their results should be accordant. The
same method is used to check the security of the transmission of the
photon 6. To prevent a dishonest party's attack and ensure the
security of the transmission of the photon 4, Alice selects randomly
Bob or Charlie to choose a random measuring basis ($Z$-basis or
$X$-basis) to measure the photon and then publish his or her
corresponding measurement result. If there is no eavesdropping, they
return to the step 1. Otherwise, the protocol is aborted.

(E4) Alice (Bob, Charlie) performs Bell basis measurement on the
photons 1 and 6 (2 and 3, 4 and 5).According to Eq. \ref{1} and
\ref{2}, we can obtain
\begin{eqnarray}
\label{3}
\ket{\phi^+_{12}}\otimes\ket{\phi^+_{34}}\otimes\ket{\phi^+_{56}}=
\frac{1}{4}(\ket{\phi^+_{16}}\ket{\phi^+_{23}}\ket{\phi^+_{45}}+\ket{\phi^+_{16}}\ket{\phi^-_{23}}\ket{\phi^-_{45}}\nonumber\\
+\ket{\phi^+_{16}}\ket{\psi^+_{23}}\ket{\psi^+_{45}}+\ket{\phi^+_{16}}\ket{\psi^-_{23}}\ket{\psi^-_{45}}\nonumber\\
+\ket{\phi^-_{16}}\ket{\phi^+_{23}}\ket{\phi^-_{45}}+\ket{\phi^-_{16}}\ket{\phi^-_{23}}\ket{\phi^+_{45}}\nonumber\\
+\ket{\phi^-_{16}}\ket{\psi^-_{23}}\ket{\psi^+_{45}}+\ket{\phi^-_{16}}\ket{\psi^+_{23}}\ket{\psi^-_{45}}\nonumber\\
+\ket{\psi^+_{16}}\ket{\phi^+_{23}}\ket{\psi^+_{45}}-\ket{\psi^+_{16}}\ket{\phi^-_{23}}\ket{\psi^-_{45}}\nonumber\\
+\ket{\psi^+_{16}}\ket{\psi^+_{23}}\ket{\phi^+_{45}}-\ket{\psi^+_{16}}\ket{\psi^-_{23}}\ket{\phi^-_{45}}\nonumber\\
+\ket{\psi^-_{16}}\ket{\phi^-_{23}}\ket{\psi^+_{45}}-\ket{\psi^-_{16}}\ket{\phi^+_{23}}\ket{\psi^-_{45}}\nonumber\\
+\ket{\psi^-_{16}}\ket{\psi^-_{23}}\ket{\phi^+_{45}}-\ket{\psi^-_{16}}\ket{\psi^+_{23}}\ket{\phi^-_{45}}).\nonumber\\
\end{eqnarray}
After the three-party's Bell basis measurements, the state of the
photons 1, 2, 3, 4, 5, 6 collapses to one of the sixteen states in
the Eq. \ref{3} with probability 1/16. Alice can then share a random
key with Bob and Charlie, as illustrate in Table 2.Suppose Bob's
result is \ket{\phi^+_{23}} and Charlie's result is
\ket{\phi^-_{45}}. If Bob collaborates with Charlie, they can deduce
that Alice's result is \ket{\phi^-_{16}} according to Eq. \ref{3}.
Thus Alice shares two-bit secret ``01'' with Bob and Charlie.

\begin{table}[h]
\caption{The establishment of sharing secret key }\label{Tab:one}
  \centering
    \begin{tabular}[b]{|c|c|c| c|} \hline
     Alice's result & \{Bob's result, Charlie's result\} & the sharing key\\ \hline
      \ \ket{\phi^+_{16}} & \{\ket{\phi^+_{23}}, \ket{\phi^+_{45}}\} \{\ket{\phi^-_{23}}, \ket{\phi^-_{45}}\} & 00\\
       & \{\ket{\psi^+_{23}}, \ket{\psi^+_{45}}\} \{\ket{\psi^-_{23}}, \ket{\psi^-_{45}}\} & \\ \hline
       \ \ket{\phi^-_{16}} & \{\ket{\phi^+_{23}}, \ket{\phi^-_{45}}\} \{\ket{\phi^-_{23}}, \ket{\phi^+_{45}}\} & 01\\
       & \{\ket{\psi^-_{23}}, \ket{\psi^+_{45}}\} \{\ket{\psi^+_{23}}, \ket{\psi^-_{45}}\} & \\ \hline
        \ \ket{\psi^+_{16}} & \{\ket{\phi^+_{23}}, \ket{\psi^+_{45}}\} \{\ket{\phi^-_{23}}, \ket{\psi^-_{45}}\} & 10\\
       & \{\ket{\psi^+_{23}}, \ket{\phi^+_{45}}\} \{\ket{\psi^-_{23}}, \ket{\phi^-_{45}}\} & \\ \hline
       \ \ket{\psi^-_{16}} & \{\ket{\phi^-_{23}}, \ket{\psi^+_{45}}\} \{\ket{\phi^+_{23}}, \ket{\psi^-_{45}}\} & 11\\
       & \{\ket{\psi^-_{23}}, \ket{\phi^+_{45}}\} \{\ket{\psi^+_{23}}, \ket{\phi^-_{45}}\} & \\ \hline
        \end{tabular}
\end{table}

We then analyze the security of the three-party QSS protocol. Each
of the communication parties transmits only one photon for each EPR
pair. The communication parties selects randomly one of the two
measuring basis ($Z$-basis and $X$-basis) to check eavesdropping.
This method for eavesdropping check is similar to that of BBM92
protocol, which is proved to be unconditionally secure. As long as
the security of the transmission of the photons 2, 4, 6 is ensured,
the present protocol is secure. The attack of Eve or a dishonest
parties will be detected in the checking mode.

The three-party QSS protocol can be easily generalized to a
multiparty one. Suppose Alice want to share a random secret key with
Bob, Charlie, Dick, $\cdots$, York and Zach. Each of the
communication parties prepares an EPR pair in the state \ket{\phi^+}
and sends one photon of the EPR pair to the next party. That is to
say Alice send one photon to Bob, Bob send one photon to Charlie,
$\cdots$, York sends one photon to Zach and Zach send one to Alice.
Similar to the three-party QSS protocol, Alice chooses checking mode
and encoding mode with probability $p$ and $1-p$, respectively. In
the checking mode, the communication parties utilizes random
$Z$-basis and $X$-basis measurement to ensure the security of the
transmitting photons. In the encoding mode, each of the
communication parties performs Bell basis measurement on their two
photons. Thus if only Bob, Charlie, $\cdots$, York and Zach
collaborate, they can share a random key with Alice. The details of
the multiparty QSS protocol is very similar to that of the
three-party one. And the security for the multiparty QSS protocol is
the same as that for three-party one.

So far we have presented a QSDC protocol and a multiparty QSS
protocol using entanglement swapping. Both the protocols utilizes
EPR pairs to achieve secure information exchange. The communication
parties transmit only one photon for each EPR pair in the two
protocols and utilize random $Z$-basis or $X$-basis measurement to
ensure the security of the transmitting photon. Different to Ref.
\cite{zhangman2}, both the parties prepares a batch of EPR pairs in
our QSDC protocol. We also point out that the sender Alice can
encode her secret message on the EPR pairs without ensuring the
security of the quantum channel firstly. In our multiparty QSS
protocol, different to Ref. \cite{zhangman}, the parties share a
random secret key without performing local unitary operations, which
simplifies the protocol.



\begin{acknowledgments}
This work is supported by the National Natural Science Foundation of
China under Grant No. 60472032.
\end{acknowledgments}

%
%

%
%
\end{document}